\documentclass[12pt,twocolumn]{article}
\usepackage{graphicx}
\usepackage{color}
\usepackage{hyperref}
\usepackage{caption}
\usepackage{setspace}
\usepackage[margin=1in]{geometry}
\usepackage{ulem}
\usepackage[T1]{fontenc}
\usepackage{amsmath}
\usepackage[utf8]{inputenc}
\usepackage{hyperref}
\usepackage{epstopdf}
\onecolumn
\begin{document}
\begin{center}
{\Large {\bf Phase Transition of 3D Heisenberg Magnet in Presence of Random-Bond Disorder }}
\end{center}
\vskip 0.5cm
\begin{center}
{\it Nepal Banerjee}\\
{\it Department of Physics,University of Seoul,South Korea}
\vskip 0.2 cm
{\bf {Email}}: nb.uos1989@gmail.com
\end{center}
\vskip 0.5cm
\begin{abstract}
Here we have simulated the random-bond type quenched disorder in 3D Heisenberg magnet.Here we have used classical Monte-Carlo simulation with Heisenberg spin and use 3D simple cubic lattice for this simulation.Here we use Metropolis single spin flipping algorithm.
\end{abstract}
\section{Introduction}
In last few decades a tremendious progress has been made in the direction of low dimensional van der Waals magnet\cite{burch2018magnetism,gong2017discovery,sinova,park2016opportunities,jiang2021recent}.Several exotic phase has emerged during the search of new material and their intrinsic properties\cite{chit2,cri3_1,KB}.However the disorder induced van der waals magnet remain still very less explore area of research\cite{aharony3,chandan,ryan1992recent,bhatt, rhodes2019disorder}.It is expected that in presence of different type of disorder this facinating magnetic material will show more exotic correlated phase.In this brief report we will summarize different type of quenched disorder and after that we will briefly present our study on  phase transition of 3D Heisenberg magnet in presence of random-bond disorder.We know that there are mostly two type of disorder and they are annealed and quenched type disorder.Quenched disorder is a static disorder and it never changes with time and annealed disorder is time dependent disorder which changes with time.Here we are mostly concerned about quenched type of disorder.This quenched disorder is a most difficult type of disorder to deal with using conventional statistical mechanics.The replica trick method made few success and able to explain the several complex phase of matter in presence of quenched type of disorder.But in several cases it also fail because of its drastic approximation.For spin system two types of quenched disorder we mostly talk about and they are random bond or random-mass disorder and another is random-field disorder.In case of random-bond disorder the static spin moments are interacting with its neighbour spin with a random exchange interaction.In a seminal work Harris pointed out the effect of this type disorder in the phase transition.He proposed a wonderful idea that if the exponent of correlation length and dimension of the system follow the criterion like $d \nu > 2 $ then this type of random-bond disorder is average out and phase transition is not going to affect in presence of disorder but if the $ d\nu < 2 $ then small disorder is going to affect the phase transition and clean critical point become unstable and sharp phase transition is impossible for that case.This is known as Harris criterion \cite{Haris}.Similarly there are another huristic argument by Imry-Ma for random-field disorder where they have shown that when a weak random field can effect the phase transition and that postulate known as Imry-Ma argument in the case of random-field disorder\cite{Imry,vojta2019disorder}.we can understand this Imry-Ma argument with a simple 3D Ising model with random field like
\begin{eqnarray}
H=-J\sum_{<i,j>} S_iS_j -\sum_{i} h_i S_i
\end{eqnarray}
Here $h_i$ are the independent random variables with zero mean and varience $[h_ih_j]_{dis}= W \delta_{i,j}$.If $W >> J^2$ then all spin gain more energy after alligning with the local field and long range feromagnetic order is impossible.But when the random filed is weak $W << J^2$ then all spin will prefer to allign with their neighbour rather than external field.Imry and Ma has given a beautiful argument to analysis the stability of this ferromagnetic order against the domain wall formation.We can eastimate this domain wall energy as  $\Delta E_{DW} = J L^{d-1}$.The energy gain because of aligning the uniform spin-up domain with the random filed is simply the sum over the random field values in the domains $\Delta E_{RF}=-\sum_i h_i$ .Using the central limit theorem we can say that this energy is propotional to $ |E_{RF}|\sim W^{1/2} L^{d/2}$.Imry and Ma observe that the uniform ferromagnetic state  will be stable against the domain formation if $|\Delta E_{RF}| < \Delta E_{DW}$.This condition is transform into $W^{1/2} L^{d/2} < J L^{d-1}$.So in case of weak random field $W << J^2$ we can say that domain formation is unfaviourable and ferromagnetic state is most stable phase for L >2.Here we have introduced a new type of random-bond disorder for 3D Heisenberg spin system.In this spin model each spin is interacting with its neighbour spin with a random exchange interaction which lies between a particular range between r to 1,where the value of r is 0.0-0.8.Here we have increased the notion of disorder after introducing the $J(A \rightarrow B) \neq J(B \rightarrow A)$.Here A and B are two nearest-neighbour lattice site.Here we have considered isotropic Heisenberg type of spin.Here we have presented our all studies in a following way.First we have discussed about the model Hamiltonian and then we have discussed the simulation methodology and results and after that we have summarize our all results as a discussion and conclusion.
\section{Model Hamiltonian}
Here we are describing our model Hamiltonian,which is a random-bond disorder induced Heisenberg model and this 3D lattice is a simple cubic lattice.Here we consider that the $J(A \rightarrow B) \neq J(B \rightarrow A)$,where A and B indicating two different nearest-neighbour lattice site.Here we consider the isotropic spin model.Here we are saying this model isotropic because the random value taken by $J_x,J_y,J_z $ are lies between same range between 0.0 to 1.0.So the mean value of $J_x ,J_y $ and $J_z$ are same.
\begin{eqnarray}
H=-\sum_{<i,j>} J_{ij} \vec S_i .\vec S_j
\end{eqnarray}
Here $\vec S_i=(S_i^x,S_i^y ,S_i^z)$ represent classical three component spin with unit magnitude sitting at each lattice point i and $J_{i,j}$ is random exchange interaction through which two nearest-neighbour spin interacting.Here we have introduced a spin model where each spin interacting with its neighbour spin with a specific random exchange interaction which lies between a particular range of value from 0 to 1.

\section{Simulation Methodology and Results}
Here we have used the classical Monte-Carlo simulation technique for simulation of disorder induced 3D heisenberg magnet.Here we have simulated the Heisenberg type of spin at each site of 3D lattice grid which have three component and we have used the following formula for this simulation\cite{banerjee2023critical,olivia,mixed,bekhechi2004chiral,beath}.
\begin{eqnarray}
S_x &=&|S|\sin(\theta) \cos(\phi)\\
S_y &=&|S|\sin(\theta) \sin(\phi)\\
S_z &=&|S|\cos(\theta)  
\end{eqnarray}
Here we have chosen the  magnitude of spin $|S|=1$ and we have simulated the different component of spin after choosing the $\theta$ and $\phi$  randomly where $\phi $ vary from 0 to $2\pi$ and  $\theta$ vary from 0 to $\pi$ respectively. 
\begin{figure}[htp]
\centering
\includegraphics[scale=0.40]{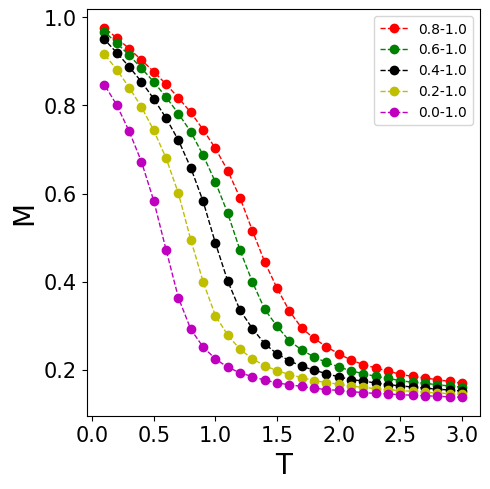}
\includegraphics[scale=0.40]{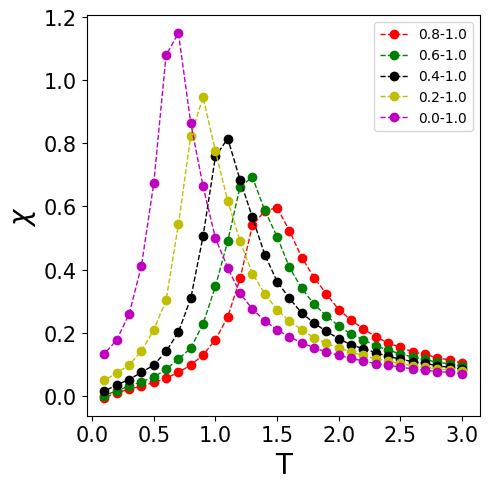}
\includegraphics[scale=0.40]{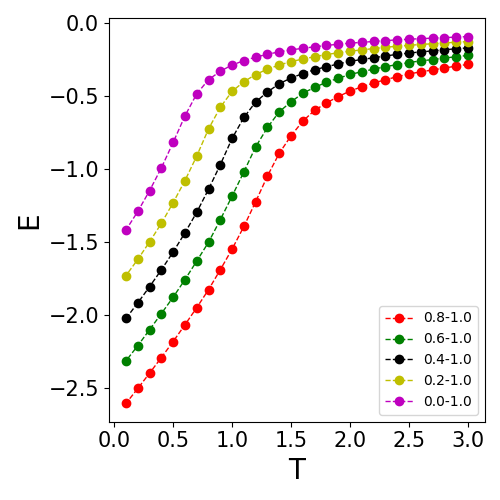}
\caption{Here we have presented the M,$\chi$,E with T.Here we have presented the results for $4 \times 4 \times 4$ lattice size.Here we have consider the value of |S|=1 and J vary randomly from 0.0-1.0,0.2-1.0, 0.4-1.0,0.6-1.0 ,0.8-1.0.}
\label{}
\end{figure}
Here we have introduced the random-bond type disorder and simultaneously simulating the exchange-interaction through which simulated Heisenberg spin will interact with its nearest-neighbour spin.In our simulation $J(A \rightarrow B ) \neq J( B \rightarrow A )$.Here A and B are two nearest neighbour.In this simulation we have used single spin flipping Metropolis algorithm and we have used $10^5$ MC steps for creating equivalent ensembles and among them we only consider $5 \times 10 ^4$ ensembles for taking the average of different thermodynamic quantity.Here we have calculated diffrent thermodynamic quantity like spontaneous magnetization(M),Susceptibility($\chi$) and energy(E) at different T.We have used the following formula for the calculation of those thermodynamic quantity.
\begin{eqnarray}
M &=&\sqrt{m_x^2 +m_y^2 +m_z^2}\\
{\chi}&=& L^3(<M^2>-<M>^2)
\end{eqnarray} 
Here $m_x=\sum S_x/L^3$,$m_y=\sum S_y/L^3$ and $m_z=\sum S_z /L^3$.Here we have used $4 \times 4 \times 4 $ lattice grid for this simulation.Here we have noticed interesting behaviour of M at different range of exchange interaction strength which is lies between 0.0-1.0,0.2-1.0,0.4-1.0,0.6-1.0,0.8-1.0.Here we have notice a significant change of scaling behaviour of M at different range of exchange interaction.Here we are noticing that the transition temperature is decreasing because of bond diluation and that is reavling at the behaviour of M  and $\chi$ with T.Here we also notice the significant change of susceptibility($\chi$) with T and we notice that the scaling behaviour of $\chi$ is changing with T.We have noticed that the peak of $\chi$ is showing more sharp peak when we increasing the range of random-bond disorder and Tc is shifting towards lower temperature.Here we have observed a ferromagnetic ground state for all the realization of random-bond disorder.We also cheak after introducing anti-ferromagnetic type random-bond disorder but small anti-ferromagnetic exchange distroy the order state of that system.

\section{Discussion and conclusion}
Here we have discussed about different type of quenched disorder and we have simulated a new type of random-bond disorder.Here we have used only one lattice size($4\times 4 \times 4$ ) just to show the basic simulation results and it can be extended easily for bigger lattice size.Here we have use classical Monte-Carlo simulation for simulating the phase transition and use Heisenberg type of spin.Here we have used a simple cubic lattice for that simulation.We observe a significant effect of random-bond disorder during the phase transition.This random-bond disorder and its dilution effect not only reduced the $T_c$ but also change scaling behaviour of the phase transition and we can clearly see that result from our simulation. 
\section{Acknowledgement}
N.B  is greatly acknowledging University of Seoul(UOS) for the funding from SAMSUNG and NRF project at initial stage of this work and we are greatly acknowledging IIT kanpur for giving visiting scholar position and providing generous research facility during the visit.Author is acknowledging Prof.P.K.Mukherjee,Prof.M.Acharyya, Prof.T.Das,Prof.J.Jung for several discussion about different aspect of the phase transition and critical phenomena over the years.Author is thankful to Prof.Chandan Dasgupta and Prof.Sumilan Banerjee for several lecture on disorder system during the course at IISc,Bangalore.Also author is thankful to Prof.Vojta,Prof.Altland for several interesting lecture during the conference at IMSc,Chennai.Author is greatly thankful to Prof.Allan H MacDonald for his fruitful collaboration with our group and several help during the progress.

\end{document}